\documentclass{iau}
\usepackage{graphicx}
\usepackage{natbib}
\usepackage{aas_macros}
\usepackage{amsmath}
\usepackage{amsfonts}
\usepackage{amssymb}

\title{Subsurface Flows Associated with Formation and Flaring Activity of Solar Active Regions}
\shorttitle{Subsurface Flows Associated with Solar Active Regions}

\author[Alexander G. Kosovichev \& Viacheslav M. Sadykov]   
{Alexander G. Kosovichev$^{1,2}$ and Viacheslav M. Sadykov$^{3}$}

\affiliation{$^1$New Jersey Institute of Technology, Newark, NJ 07102, U.S.A. \\ email: {\tt alexander.g.kosovichev@njit.edu} \\
	$^2$NASA Ames Research Center, Moffett Field, CA 95035, U.S.A.\\
	$^3$ Physics \& Astronomy Department, Georgia State University, Atlanta, GA 30303, U.S.A.}

\pubyear{2024}
\volume{365}  
\pagerange{119--126}
\jname{Dynamics of Solar and Stellar
Convection Zones and Atmospheres}
\editors{A.V. Getling \& L.L. Kitchatinov, eds.}

\begin{document}
	
\maketitle

\begin{abstract}
We investigate the evolution of subsurface flows during the emergence and the active phase of sunspot regions using the time-distance helioseismology analysis of the full-disk Dopplergrams from the Helioseismic and Magnetic Imager (HMI) onboard the Solar Dynamics Observatory (SDO). We present an analysis of emerging active regions of various types, including delta-type active regions and regions with the reverse polarity order (`anti-Hale active regions'). The results reveal strong vortical and shearing flows during the emergence of magnetic flux, as well as the process of formation of large-scale converging flow patterns around developing active regions, predominantly in the top 6 Mm deep layers of the convection zone. Our analysis revealed a significant correlation between the flow divergence and helicity in the active regions with their flaring activity,  indicating that measuring characteristics of subsurface flows can contribute to flare forecasting.

\end{abstract}

\firstsection 

\section{Introduction}
Most manifestations of solar activity are associated with magnetic sunspot regions emerging from the Sun's interior. The bipolar structure of active regions indicates that they originate from the toroidal magnetic flux generated by solar dynamo, but the physical processes of the flux emergence and formation of active regions and sunspots are not understood. In addition, the origin of complex magnetic configurations, such as delta-type sunspots with strong magnetic field gradients across the polarity inversion line, leading to the impulsive release of magnetic energy in the form of solar flares and coronal mass ejection, remains unclear. 

Observations of photospheric magnetic fields show that the formation of active regions starts from the emergence of fragmented magnetic patches that are subsequently collected in compact, stable magnetic structures, observed as pores and sunspots. However, it remains unclear whether this process reflects `monolithic' magnetic structures formed inside the Sun prior to the flux emergence or magnetic self-organization controlled by the near-surface plasma flows associated with emerging magnetic structures.

The process of magnetic self-organization on the Sun was demonstrated in 3D radiative MHD simulations by \citet{Kitiashvili2010}. In these simulations, the initial uniformly distributed 100~G field collapsed into a pore-like structure with the magnetic field strength reaching 1.5~kG on the surface and $\sim 6$~kG below. This structure is maintained by self-forming converging downdrafts confirming the idea initially suggested by \citet{Parker1955}. A similar spontaneous formation of a pore-like structure was described by \citet{Stein2012} who simulated the emergence of initially a uniform horizontal magnetic field and confirmed that the self-organization is an intrinsic feature of magnetic fields emerging on the solar surface. 
 
However, attempts to model the spontaneous formation of stable sunspot-like structures have not been successful. \citet{Rempel2014} modeled the flux emergence driven by advecting a semi-torus-like structure of the magnetic field at the bottom boundary. This model showed the formation of a bipolar magnetic structure at the surface. However, it quickly decayed due to the turbulent diffusion. \citet{Hotta2020} simulated the formation of sunspot-like structures from an initially horizontal twisted magnetic flux tube located at a depth of about 35 Mm, a part of which was advected by convective upflows. It was demonstrated that the converging downdrafts are a key mechanism of the structure formation. Also, it was found that the temperature distribution and flow patterns beneath the structures resemble the helioseismology results \citep{Kosovichev2000,Zhao2001, Zhao2003}. However, like in the previous simulations of \citet{Rempel2014}, the decay of the sunspot-like structures started immediately after their formation when the downflows in these structures turned into upflows. To overcome this difficulty,  \citet{Brandenburg2016} investigated the formation of magnetic structures in a forced stratified turbulent layer and demonstrated that the effective magnetic pressure may become negative and result in magnetic field concentrations. However, these concentrations were found to be too deep and too narrow to explain the sunspots \citep{Perri2018}. 
 
It has been known for a long time that complex active regions, in which strong magnetic fields of the opposite polarity form polarity inversion lines with high magnetic field gradients, are a primary source of large solar flares and eruptions \citep{Severnyj1958,Severny1964}. Magnetic structures of this type (called delta-type sunspots) can be formed when magnetic fields with opposite polarities emerge close to each other or pushed toward each other after the emergence \citep{Zirin1987}. Such magnetic configurations have been reproduced in the simulations of \citet{Kaneko2022} who employed the flux-tube model of \citet{Hotta2020} for various distributions of upwelling convective flows. However, contrary to observations, in these simulations, the delta-type sunspot-like structures were forming more frequently than the simple beta-type sunspots. 

In general, the MHD simulations demonstrated that subsurface flows associated with the emergence and evolution of the magnetic field play a critical role in the formation, stability, and activity of sunspot regions. The observational information about the subsurface flows is provided by local helioseismology techniques: ring-diagram analysis \citep{Gough1983,Komm2005}, time-distance helioseismology \citep{Duvall1993,Kosovichev1996}, and acoustic holography \citep{Braun2000}. The subsurface flow maps obtained by the ring-diagram technique detected large-scale converging flows around developed active regions in the top 7~Mm-deep layer and diverging flows at a depth of 14~Mm \citep{Haber2004}. These flows, which may significantly affect the meridional magnetic flux transport, were confirmed by the time-distance helioseismology \citep{Zhao2004}. 

The primary goals of our study are to improve the characterization and physical understanding of the emergence and evolution of active regions and explore the potential of utilizing physical descriptors of subsurface flows for predicting periods of flaring activity. The processes by which the magnetic energy is stored, enters, and leaves active solar regions, are critically linked to the flow patterns therein. Large-scale organized flows are developed spontaneously in subsurface layers due to the emergence of magnetic flux and its interaction with the existing magnetic field of active regions. This process forms stressed magnetic configurations that trigger solar eruptions. 

In Section 1, we briefly review the time-distance helioseismology pipeline methodology implemented in the Solar Dynamics Observatory's (SDO) Joint Science Operation Center (JSOC), and the procedure for tracking the evolution of subsurface flows in emerging active regions. In Section 2, we present and discuss the evolution of subsurface flows for several active regions, as well as the relationship of various flow characteristics to the flaring activity. We present initial results on the correlation of flow characteristics with the flare productivity of active regions and Section 3. In Section 4, we discuss the outcome of this study and future perspectives. 

\section{Measurements of Subsurface Flows by Time-Distance Helioseismology}

\begin{figure}
	\begin{center}
		\includegraphics[width=0.8\linewidth]{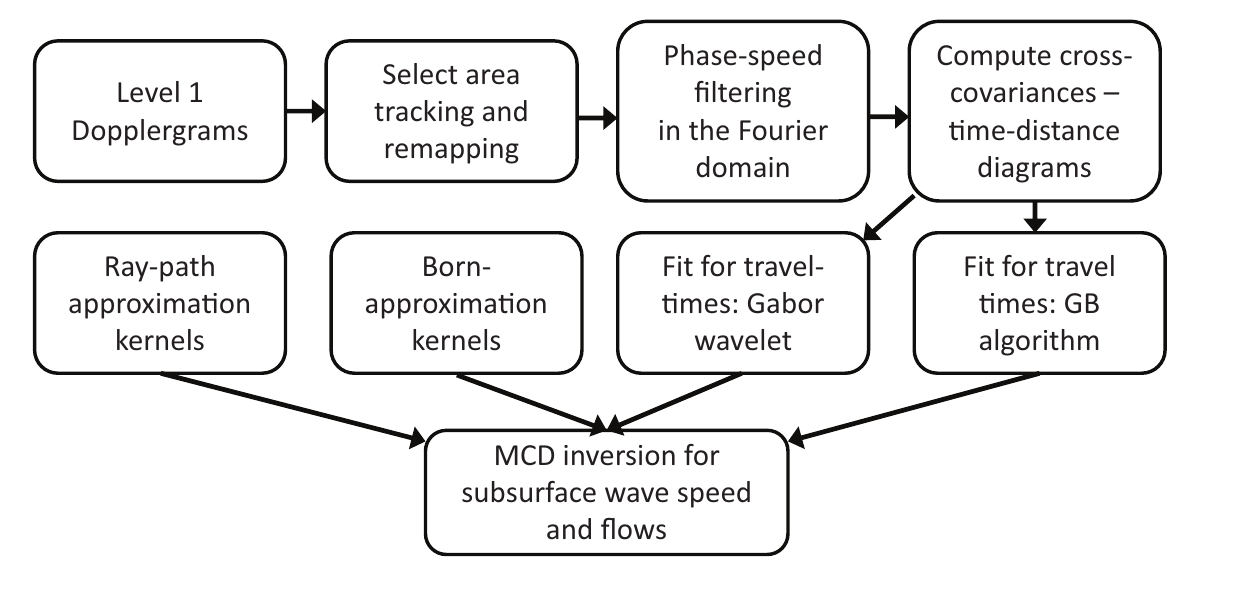} 
		\caption{
			A schematic representation of the time-distance helioseismology pipeline \citep{Zhao2012}. 
		}\label{fig1}
	\end{center}
\end{figure}

The computational pipeline (Fig.~\ref{fig1}) for studying the subsurface dynamics of active regions takes the Carrington coordinates of active regions at the central meridian from the Solar Region Summary (SRS) database, compiled by the National Oceanic and Atmospheric Administration's (NOAA) Space Weather Prediction Center (SWPC), and uses these coordinates as the central points of 30-degree areas tracked for 10 days during their passage on the solar disk. 
This setup allows us to follow the evolution of active region areas even before the magnetic flux emergence (i.e., retroactively) and after the decay. The 3D subsurface flow maps are calculated from the tracked Dopplergrams that are remapped onto the heliographic coordinates using Postel's projection (transverse cylindrical projection that preserves the distance along great circles). 
Each tracked, 8-hour-long datacube consists of 640 Dopplergrams of pixels with a spatial resolution of 0.06 degree/pixel and a 45-second time cadence. The tracked datacubes are processed through the Time-Distance Helioseismology Pipeline \citep{Zhao2012,Couvidat2012}. The output represents acoustic travel-time maps calculated with 0.12-deg sampling for the whole tracked area (512$\times$512 pixels).

The travel times are calculated for eleven annuli located at different distances from central points corresponding to 2x2-binned original Dopplergram pixels. The signals of acoustic waves traveling between the central points and the surrounding annuli are calculated from the HMI Doppler velocity measurements as the corresponding cross-covariance functions. The cross-covariances are computed in the Fourier space, and phase-space filters are applied to isolate the signals corresponding to each of the travel distances.

$$
\psi(\tau, \Delta)=\int_{0}^{T} f(t, \boldsymbol{r}) f^{*}(t+\tau, \boldsymbol{r}+\Delta) d t
$$
The travel times are calculated by two different methods: 1) the Gabor wavelet fitting (Kosovichev and Duvall, 1997)

$$
G\left(A, \omega_{0}, \delta \omega, \tau_{\mathrm{p}}, \tau_{\mathrm{g}} ; t\right)=A \cos \left\{\omega_{0}\left(t-\tau_{\mathrm{p}}\right)\right\} \exp \left(-\frac{\delta \omega^{2}}{4}\left(t-\tau_{\mathrm{g}}\right)^{2}\right)
$$
and 2) a cross-correlation with a reference (Gizon \& Birch 2002).

The travel times are used to infer the 3D maps of subsurface flows by solving an inverse problem. It is formulated as a set of linear integral equations whose kernels are calculated by using either the ray-path theory or the Born approximation:

$$
\delta \tau\left(x_{1}, x_{2}\right)=\int_{V} \mathrm{~d} \boldsymbol{r} \boldsymbol{K}\left(\boldsymbol{r} ; \boldsymbol{x}_{1}, \boldsymbol{x}_{2}\right) \cdot \boldsymbol{v}(\boldsymbol{r})
$$

The system of integral equations can be discretized and formulated as the set of linear equations for grid points $(i, j, k)$ for each surface point $(\lambda, \mu)$ and travel distance, $\nu$:

$$
\delta \tau_{l}^{\lambda \mu \nu}=\sum_{i j k, l} A_{i j k, l}^{\lambda \mu \nu} \tilde{v}_{l}^{i j k}. 
$$

Here, $\quad \tilde{v}_{l}^{i j k}=\left[v_{x}^{i j k}, v_{y}^{i j k}, v_{z}^{i j k}\right] / c^{k} $ is the vector of velocity at the horizontal grid points $i, j$, and vertical grid points, $k$, relative to the sound speed at these points, calculated from the standard solar model. Regularized solutions are determined by the Multi-Channel Deconvolution (MCD) method \citep{Couvidat2005}, and the regularization parameters are chosen to suppress noise and represent a smooth solution.

We used the SDO/HMI time-distance helioseismology pipeline (Fig.~\ref{fig1}) to infer 3D subsurface flow maps during the emergence and evolution of Active Regions. The travel times are used for the reconstruction of subsurface flows in 8 subsurface layers in the depth ranges: 0-1, 1-3, 3-5, 5-7, 7-10, 10-13, 13-17, and 17-21 Mm, and with the horizontal spatial sampling of 0.12 degrees (1.5 Mm). The horizontal and vertical resolution is determined by the averaging kernels calculated in the inversion procedures. The characteristic width of the averaging kernels is roughly proportional to the wavelength of acoustic waves and increases within this range of depth from $\sim 2$ to $10$~Mm \citep{Couvidat2005,Parchevsky2014}. Therefore, the flow maps obtained from the helioseismic inversions represent smoothed versions of the actual subsurface flows.  Combinations of two travel-time measurement techniques and two types of sensitivity kernels provided four flow maps for each time interval. The flow patterns in these maps are very similar. Some differences in the flow amplitude (mainly in the top layers) can be attributed to differences in the sensitivity and averaging kernels. In this paper, we present the flow maps obtained by using the Gabor-wavelet technique for travel-time measurements and the Born-approximation kernels.

The primary source of noise is due to the random excitation of solar oscillations, the so-called ``realization noise.'' In the flow maps obtained for a particular depth, it causes random variations with the characteristic scale corresponding to the wavelength of acoustic waves with the turning point located at this depth. These variations are significantly weaker than the signal from convective flows for the depths up to $8$~Mm but become greater than the convective velocities in the deeper layers. However, the flows associated with active regions are significantly stronger than the quiet-Sun convection in these layers. 

The helioseismic inversions for all active regions presented in this paper show large-scale divergent flows at depths greater than 6 Mm, similar to the flows around isolated sunspots \citep{Kosovichev2000,Zhao2003,Kosovichev2011}. Therefore, we focus on the dynamics of the shallower 4~Mm deep layers which seems to play a key role in the formation and dynamics of sunspots and active regions. We primarily discuss the horizontal velocity maps because the inversions for the vertical velocity may have systematic errors due to a potential cross-talk with the horizontal flow components, particularly in the regions with regular horizontal velocity patterns and small vertical velocities, e.g. in the near-surface layers of supergranulation cells. However, the testing with artificial data showed that the inferences of stronger vertical velocities in the deeper layers are not affected by the cross-talk \citep{Zhao2003a}. 

\section{Subsurface Dynamics During Emergence and Evolution of Active Regions}

In this Section, we present four examples of emerging active regions of various types:
\begin{itemize}
\item AR~11158, a delta-type region formed by the interaction of two emerging bipolar magnetic regions,
\item AR~12673, a delta-type region formed by the collision of a large sunspot with emerging bipolar magnetic flux,
\item AR~12282, an anti-Hale active region formed in the vicinity of a large sunspot,
\item AR~13006: a small delta-type anti-Hale active region that produced an X-class flare,
\item AR~13179, a beta-type active region formed by the continuous emergence of small-scale bipolar magnetic elements.
\end{itemize}

\begin{figure}
	\begin{center}
		\includegraphics[width=\linewidth]{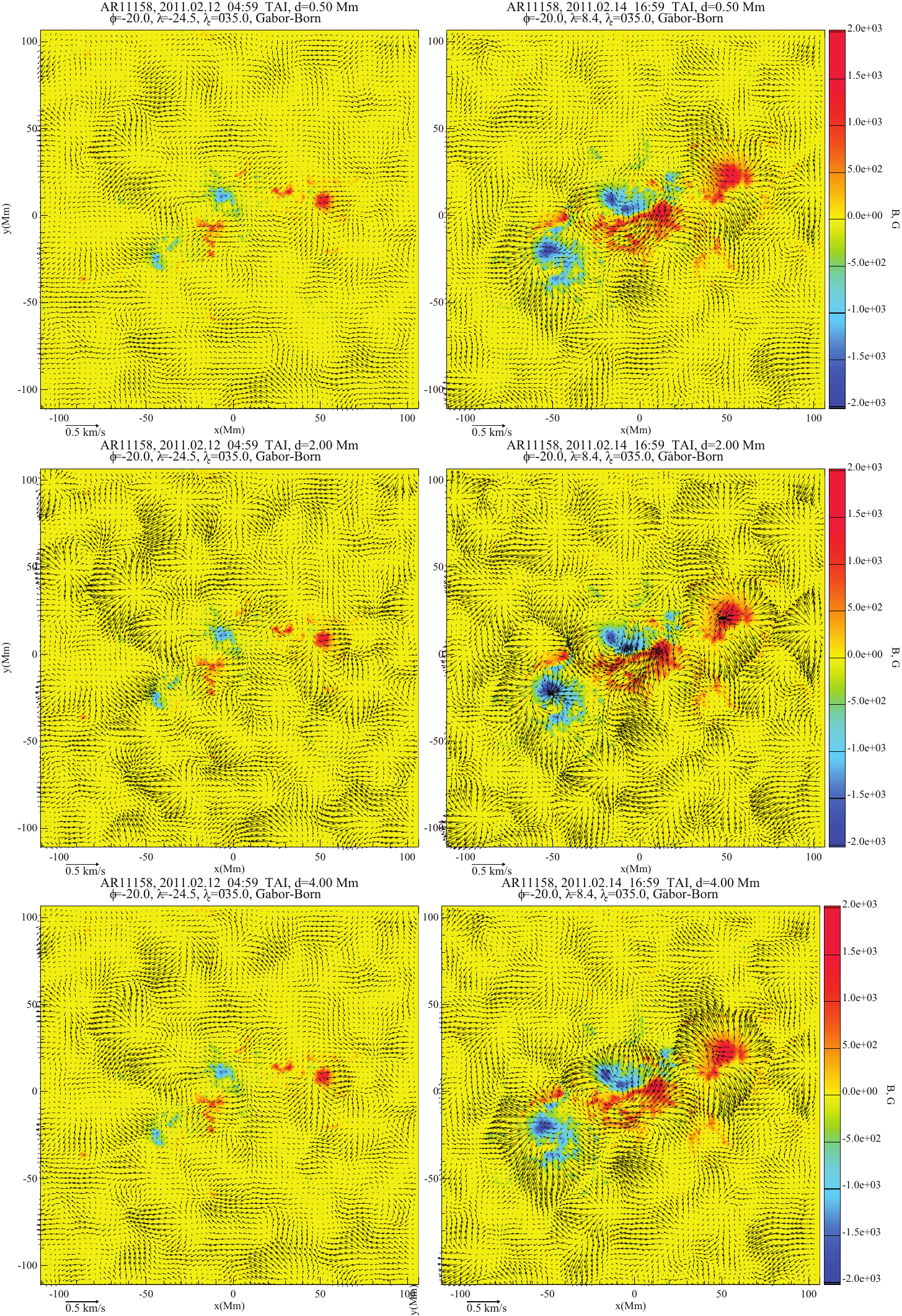} 
		\caption{Flow maps of AR 11158 during the emergence (left column) and before the X2.2 flare (right) at three depth levels centered at 0.5, 2, and 4~Mm below the solar surface (from top to bottom). The background images show the corresponding surface magnetograms.}
		\label{fig2}
	\end{center}
\end{figure}

While these examples do not cover all active region types, they provide interesting examples of the formation of flare-productive delta-type regions, relatively simple regular beta-type regions, and the mysterious `anti-Hale' regions with the polarity order violating Hale's polarity law \citep{Hale1919}. 

In Figures \ref{fig2}-\ref{fig6}, we present the flow maps for each of these active regions for two moments of time, during the emergence and the developed state, for three top layers at the depths, 0-1~Mm, 1-3~Mm, and 3-5~Mm. We use the corresponding central locations 0.5, 2, and 4~Mm for the identification of these layers. The corresponding surface line-of-sight magnetograms are shown in the color images. The magnetic field of the positive polarity is shown in red, and the negative polarity is shown in blue.  The titles at the top of each panel in these figures show the midtime of the flow maps, obtained from the 8-hour Dopplergram series, the depth $d$, the latitude $\phi$, longitude relative to the central meridian $\lambda$, and the Carrington longitude $\lambda_c$ of the center of the images at $x=y=0$. For the time-distance analysis, the areas on the solar disk are remapped onto the heliographic coordinates, using Postel's projection centered at the center of these images. For convenience, the heliographic coordinates are converted in the linear distances along the surface, and shown in megameters along the $x$ and $y$ axes. The horizontal velocity vectors are shown by arrows, the arrow scale is shown at the bottom of each panel (kept the same for all active regions). The label Gabor-Born indicates that the travel times were determined by the technique of fitting the Gabor wavelets to the oscillation cross-covariance function \citep{Zhao2012}, and the Born-approximation kernels were used for the travel-time inversion \citep{Birch2000,Birch2004,Birch2007}.   

\subsection{AR 11158: Delta-type Region Formed by Interaction of Two Emerging Bipolar Magnetic Regions}

Active region 11158 initially emerged on 2011.02.09 as a simple bipolar magnetic region (BMR). About 20 hours later, a second bipolar structure emerged about 50 Mm from the first BMR. The leading positive polarity was rapidly moving apparently by local subsurface flows, and on 2011.02.12, it reached the trailing negative polarity of the first BMR (Fig.~\ref{fig2}, left column). During the subsequent magnetic flux emergence in both BMRs, the positive polarity of the second BMR overtook the trailing polarity region of the first BMR. The interaction of these magnetic structures created a delta-type configuration with a strong magnetic field gradient along the polarity inversion line, which resulted in X2.2 flare on 2011.02.15. The right panel in Fig.~\ref{fig2} shows the subsurface flows and the surface magnetic field on 2011.02.14, about 9 hours before the flare. The flows around the mixed polarity structure in the middle of this image reveal a large-scale vortex in the top 2 Mm layer. This vortex pushed the magnetic field of the positive polarity towards the negative polarity spot, enhancing the magnetic field gradient across the polarity inversion line. The flows converging towards the polarity inversion line are also visible in the 4~Mm layer. This indicates that the subsurface flows may play a significant role in the formation and flaring activity of the delta-type active regions. 

The flows outside the isolated sunspots are predominately diverging, while they are converging in these layers beneath the sunspots, corresponding to the previous results of \citet{Kosovichev2000,Zhao2001}. 

The relationship between the surface and subsurface flows in this active regions was studied by \citet{Liu2013} who compared flows obtained by applying the Differential Affine Velocity Estimator for Vector Magnetograms (DAVE4VM) technique with the near-surface velocity from the time-distance helioseismology measurements in the 0.5~Mm deep layer. They found general similarities between surface and subsurface flows but also significant differences in some areas. These differences may be due to rapid variations of plasma flows with depth or due to systematic differences in the measurement techniques. In particular, the DAVE4VM technique determines the flow velocity from the vector magnetic field variations by assuming that the magnetic field is frozen into plasma \citep{Schuck2008}, which is not necessarily the case because of the turbulent diffusion. On the other hand, the spatial and temporal resolutions of the time-distance technique are significantly lower than those of the DAVE4VM method.

\subsection{AR 12673: Delta-type Region Formed by Collision with Emerging Emerging Bipolar Region}
\begin{figure}
	\begin{center}
		\includegraphics[width=\linewidth]{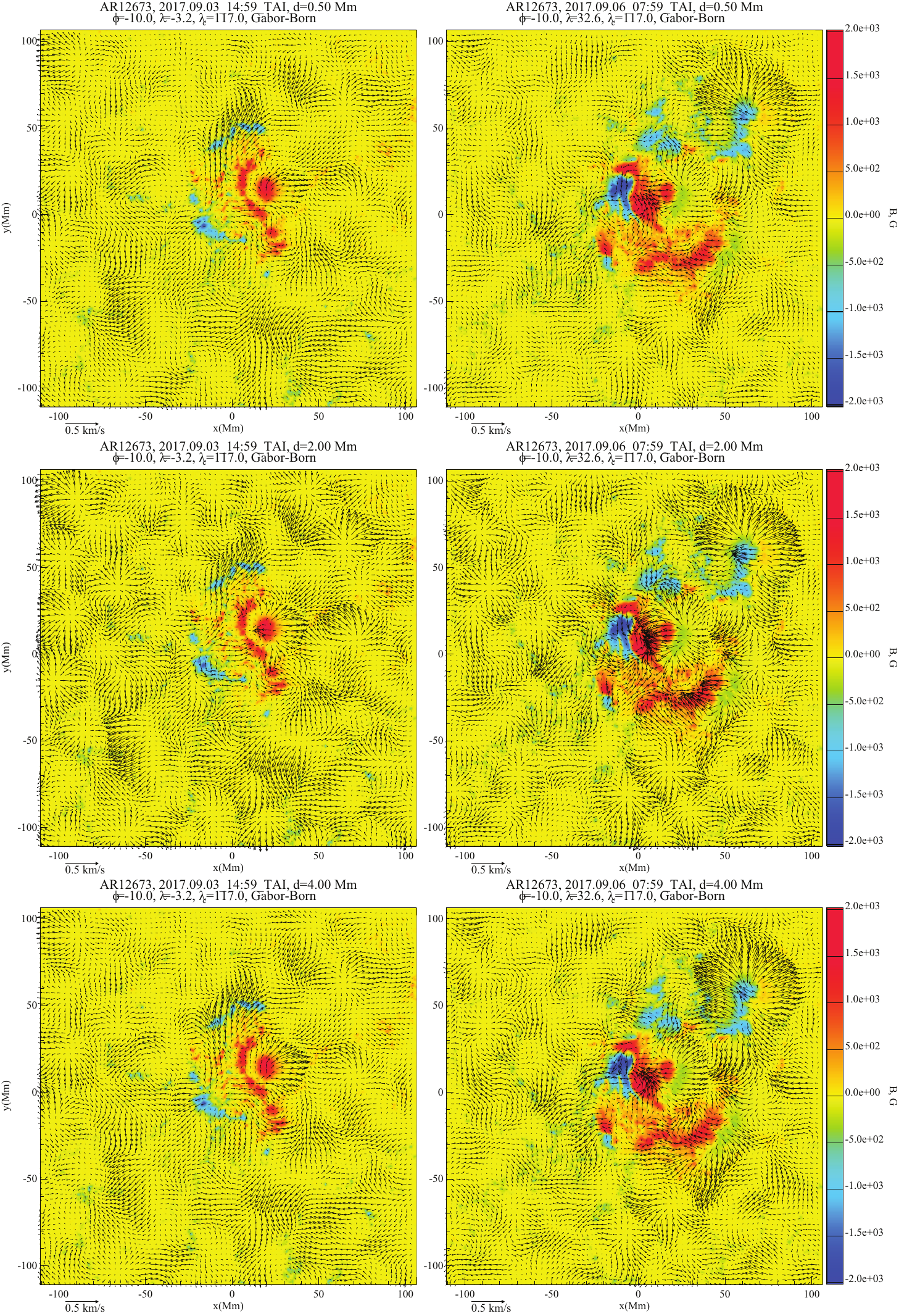} 
		\caption{Flow maps of AR 12673 during the emergence of bipolar magnetic flux in the form of fragmented ribbons near the existing positive polarity sunspot (left column) and before the X9.3 flare (right) at four depth levels centered at 0.5, 2, 4, and 6~Mm (from top to bottom). The background images show the corresponding surface magnetograms.
		}
		\label{fig3}
	\end{center}
\end{figure}

\begin{figure}
	\begin{center}
		\includegraphics[width=\linewidth]{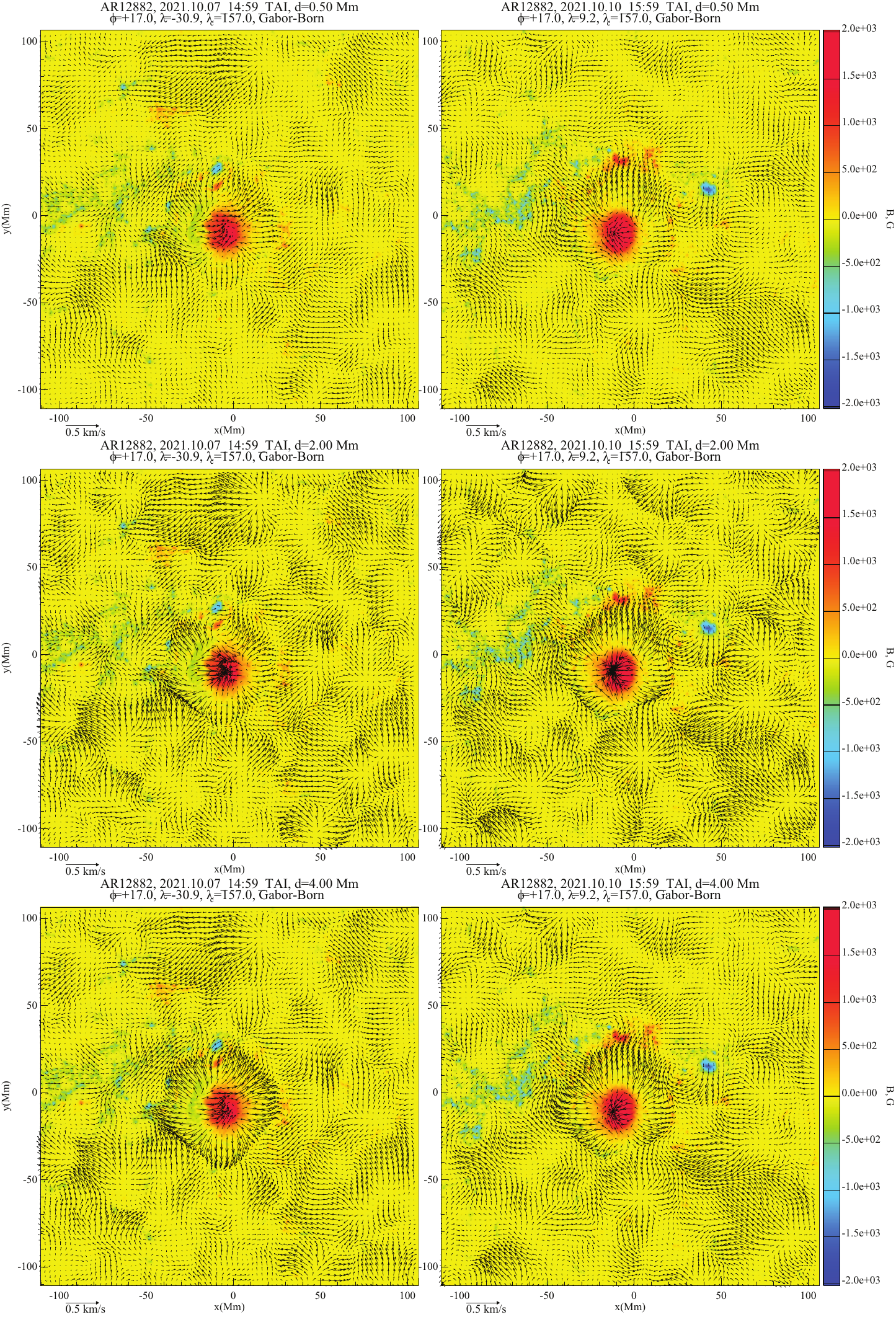} 
		\caption{Subsurface flow maps of AR 12882 during the initial emergence of a small bipolar magnetic region just above of a large positive polarity sunspot (left), and at the developed stage of an anti-Hale region (right). The background images show the corresponding surface magnetograms.	}
		\label{fig6}
	\end{center}
\end{figure}
Active region 12673 presents another interesting example of the formation and flaring activity of a delta-type region with a strong magnetic gradient across the polarity inversion line (Fig.~\ref{fig3}). The delta-type configuration of this active region was formed by the interaction of new emerging bipolar flux with the existing stable sunspot. Initially, a small bipolar region emerged on 2017.09.02 south to the sunspot, but then the flux emergence quickly expanded to the north forming ring-like structures around the sunspots by 2017.09.03 (left panel in Fig.~\ref{fig3}). The northern part of the emerging negative polarity flux was moving fast to the east and formed a stable sunspot (in the upper right corner in the right panel of Fig.~\ref{fig3}). The portion of the emerging flux, which was located south of the original sunspot, created an extended sunspot system of positive polarity. The portions of the emerging flux, which were in the latitudinal range as the original sunspot, collided with this sunspot and created a delta-type configuration with a longitudinally extended polarity inversion line. This polarity inversion line was a source of several strong flares, including the X9.3 flare on 2017.09.06, which produced a series of helioseismic events - sunquakes \citep{Sharykin2018}.  

The right panel in Fig.~\ref{fig3} shows the subphotospheric flows just before the big flare. The flow structure in this region is very complicated. We observe significant variations not only in the flow speed but also in the flow direction with the depth. The most notable feature is the converging flows towards the polarity inversion line, which are the strongest in the 2 and 4~Mm deep layers, and the large-scale vortex flow around the negative polarity in the top two layers near the center of the presented flow maps. This vortex flow converges to the polarity inversion line from the south. Curiously, the east-ward flow towards the polarity inversion line stops at about 2~Mm from the polarity inversion line. At this point, it is unclear whether it is a physical effect or a result of systematic errors (e.g. caused by the suppression of acoustic waves in strong field regions). Shearing flows in opposite directions are developed along the polarity inversion line in the top subsurface layer. However, these are not as strong as observed by feature tracking techniques on the surface.

The surface flows in this region were previously studied by \citet{Getling2019a,Liu2023,Verma2018,Wang2018} using various techniques of local correlation tracking and the DAVE4VM technique. These studies also revealed strong shearing and vortex flows and high vertical velocities in the vicinity of the polarity inversion line.  The subphotospheric flows are generally consistent with these conclusions, but there are some differences, which are likely to be attributed to differences in the spatial and temporal resolutions and to variations of the flow patterns with depth. In particular, the helioseismic maps reveal flows crossing the polarity inversion line, which the surface measurements cannot detect by the nature of these techniques.

\begin{figure}
	\begin{center}
		\includegraphics[width=\linewidth]{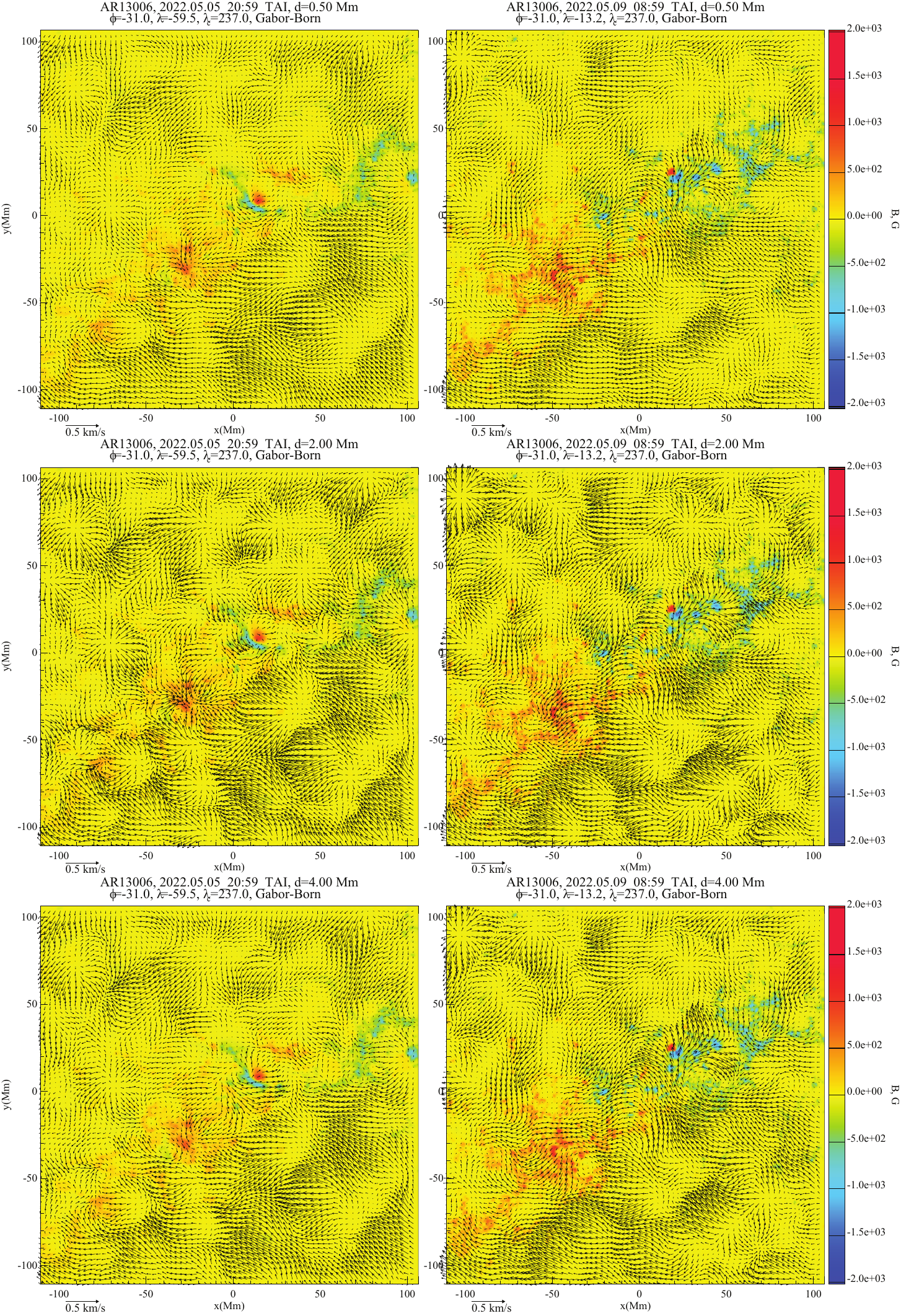} 
		\caption{Flow maps in the area of a small delta-type anti-Hale active regions AR 13006, which formed at a boundary of a large-scale diffused bipolar magnetic region (forming an inclined belt of fragmented magnetic flux). The emergence of additional negative polarity flux (blue) around the round positive polarity sunspot (red) caused an X1.2 flare on May 10, 2022.    The background images show the corresponding surface magnetograms.
		}
		\label{fig4}
	\end{center}
\end{figure}

\subsection{AR 12882: delta-type Anti-Hale Active Region}

Most active regions follow Hale's polarity law  \citep{Hale1919}, according to which the magnetic polarity of the leading sunspots is the opposite in the northern and southern and changes the polarity every 11 years. However, in about 3\% of the observed active regions, the leading sunspots have a polarity that is opposite to the majority of active regions, thus, violating the Hale law. Figure~\ref{fig6} shows subsurface flows in AR~12882, in which an anti-Hale active region was formed. The formation of the anti-Hale magnetic structure started from the emergence of a small bipolar magnetic region in the vicinity of a large sunspot, oriented in the South-North direction (Fig.~\ref{fig6}, left panels). Then, in the next three days, the bipolar region continued to grow, and the process of the magnetic flux emergence was similar to the emergence of a normal `Hale' active region, accompanied by the accumulation of small-scale magnetic elements and the polarity separation. However, the negative (blue) polarity quickly moved forward, apparently driven by local west-directed flows, and formed a small negative polarity sunspot (Fig.~\ref{fig6}, right panels). At this time, the flows around these sunspots were diverging. Curiously, the interaction of the outflows from this sunspot and the following big positive polarity sunspot generated a vortex between them. The negative polarity spot continued moving away from the positive polarity sunspot and decayed in about 4 days. The fast polarity separation may indicate that the emerging `anti-Hale' sunspot belonged to a toroidal flux system with the field direction opposite to the primary toroidal flux. Such scenarios of the formation of the anti-Hale bipolar magnetic regions were previously suggested by \citet{Stenflo2012}.

\subsection{AR 13006: delta-type Anti-Hale Active Region With X-class Flare}

AR~13006 is another interesting example of the delta-type active region that produced the X1.5 flare with a strong sunquake on 2022.05.10. This active region was formed near the boundary of the polarity reversal in a very large elongated diffuse magnetic field structure. This active region appeared in this form on the East limb; thus, its initial emergence was not observed. In this region, a small but stable positive polarity sunspot was surrounded by the magnetic field of the opposite polarity (Fig.~\ref{fig4}, left panel). Interestingly, the supergranulation flows look substantially stronger south of the elongated magnetic structure than north of it, which is apparently caused by large-scale flows directed toward this structure. This flow is particularly strong in the 2 and 4~Mm deep layers. In the deeper layers, the supergranulation flows diminish but the large-scale flow directed north-east remains. Probably, it reflects a deep circulation of a giant-cell scale, correlated with large-scale magnetic structures in the convection zone. 

As the active region evolved, the negative polarity continued to emerge on the southeast side of the positive polarity sunspot. Such emergence was accompanied by local northwest-directed flows, which pushed this flux closer to the sunspot. One of these flux-emerging events created a ring-like structure around the sunspot, forming a polarity inversion line with a strong magnetic field gradient, which was the source of the X1.5 flare and a strong sunquake \citep{Kosovichev2023}. After the flare, the positive polarity sunspot quickly disappeared. 

The right panel in Figure~\ref{fig4} shows the flows and photospheric magnetic field on 2022.05.09, a day before the flare. At that time, the negative polarity of bipolar magnetic flux elements, which were emerging in the area of the polarity inversion of the large-scale magnetic structures, were moved towards the positive polarity spot. It seems that the expansion of continuously emerging small-scale bipolar magnetic elements and their merging into larger-scale magnetic structures is a common mechanism of the active region formation and that the emerging magnetic flux has a fragmented small-scale turbulent structure, which is organized in large-scale sunspot regions in the relatively shallow near-surface layers. Flare-productive delta-type configurations can be formed during this process when the new flux emerges in the vicinity of previously emerged magnetic structures. The subphotospheric flow maps indicate that the plasma flows resulting from the interaction of emerging magnetic flux with the turbulent convection in the top 4-6~Mm deep layer, play a key role in the magnetic self-organization of magnetic field on the solar surface. 

\begin{figure}
	\begin{center}
		\includegraphics[width=\linewidth]{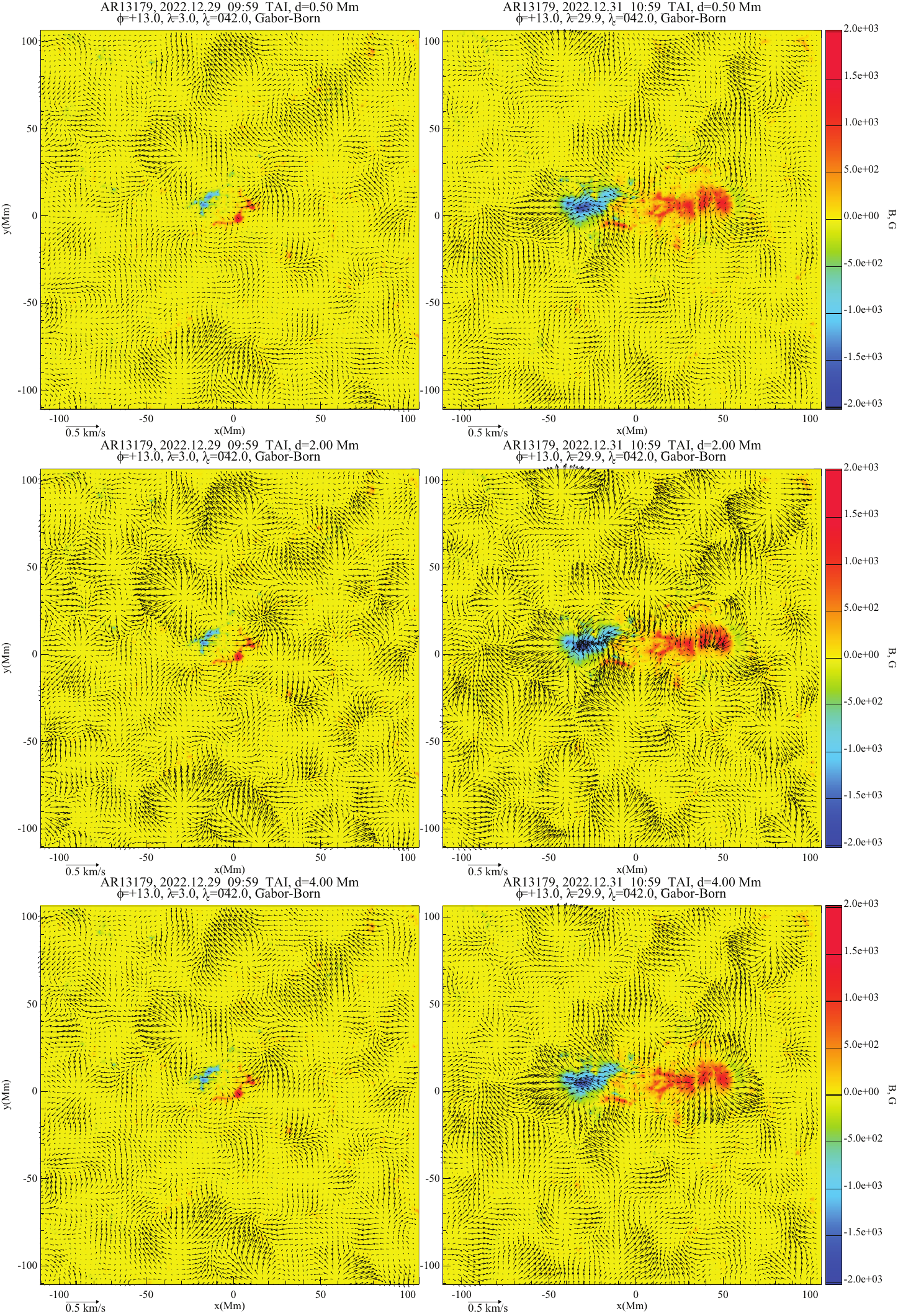} 
		\caption{Flow maps of AR 13179 during the initial emergence of bipolar magnetic flux (left) and the developed active region phase. The active region was formed through the accumulation of small-scale bipolar magnetic elements continuously emerging in the middle area of this region. The background images show the corresponding surface magnetograms.
		}
		\label{fig5}
	\end{center}
\end{figure}

\subsection{AR 13179: Formation of a $\beta$-type Active Region}

Active region 13179 provides an interesting example of the magnetic self-organization process. It started with the emergence of a compact circular-shaped bipolar region at a boundary between supergranules on 2022.12.29 (Fig. \ref{fig5}, left panel). It quickly expanded, apparently reflecting a large underlying magnetic structure. This expansion was accompanied by the flows diverging from the central area of the active region. The subsequent emergence occurred in the region separating the positive and negative polarities in the form of continuous emergence of small-scale bipolar elements. The positive and negative elements were streaming towards the forming sunspots of the corresponding polarity. The rate of magnetic flux emergence in the central area of the active region was about $2\times 10^{16}$~Mx$\,$s$^{-1}$. Thus, the total magnetic flux that emerged in the central area during the three days of the active region formation was about $5\times 10^{21}$~Mx, which is about half of the total active region magnetic flux.

At the beginning of the active region emergence, the subsurface flows corresponded to an expansion from the central area towards the growing magnetic polarities (Fig.~\ref{fig5}, left panel). When the active region developed, the inflows into the sunspot areas were observed, which presumably increased the accumulation of the magnetic flux. In addition, strong outflows extended outside the sunspot penumbras were developed predominantly in the longitudinal directions (Fig.~\ref{fig5}, right panel). Usually, the flow patterns in the sunspot areas, constituting compact converging flows beneath the sunspots in shallow $\sim 4$~Mm deep layers, and large-scale outflows in the deeper layers, are more symmetrical. In this case, the inversion results show a stream of flows through the trailing sunspot. It was more compact and better developed than the leading fragmented sunspot, which is unusual.

\section{Flow Characteristics and Flaring Activity}

The subsurface flow maps revealed shearing and converging flows around the polarity inversion lines, which are the places where energy is released in solar flares. For example, the flow maps of AR~12673 reveal shearing flows beneath the active region (Fig.~\ref{fig3}, right panels) and, in particular, in the area of the polarity inversion line, which was the source of several flares, including the X9.3 flare of Sept. 6, 2017. 

Therefore, it is important to investigate the links between the flow characteristics and the flaring activity.  For this analysis, we derived the physical descriptors to characterize the subsurface flow maps: a layer-average horizontal divergence calculated from the horizontal velocity components, a layer-average vertical component of the flow vorticity, and a proxy for the kinetic helicity defined as a product of the horizontal divergence and the vertical component of vorticity. 

We used the horizontal components of the velocity for the calculations of the flow divergence, vorticity, and helicity because the vertical component of the subsurface flows is determined with significantly larger uncertainties. Nevertheless, we use the averaged vertical velocity as a separate descriptor. The three subsurface layers closest to the surface (0-1 Mm, 1-3 Mm, and 3-5 Mm deep) are most relevant to the short-term active region evolution. 

\begin{figure}[h!]
	\begin{center}
		\includegraphics[width=0.96\linewidth]{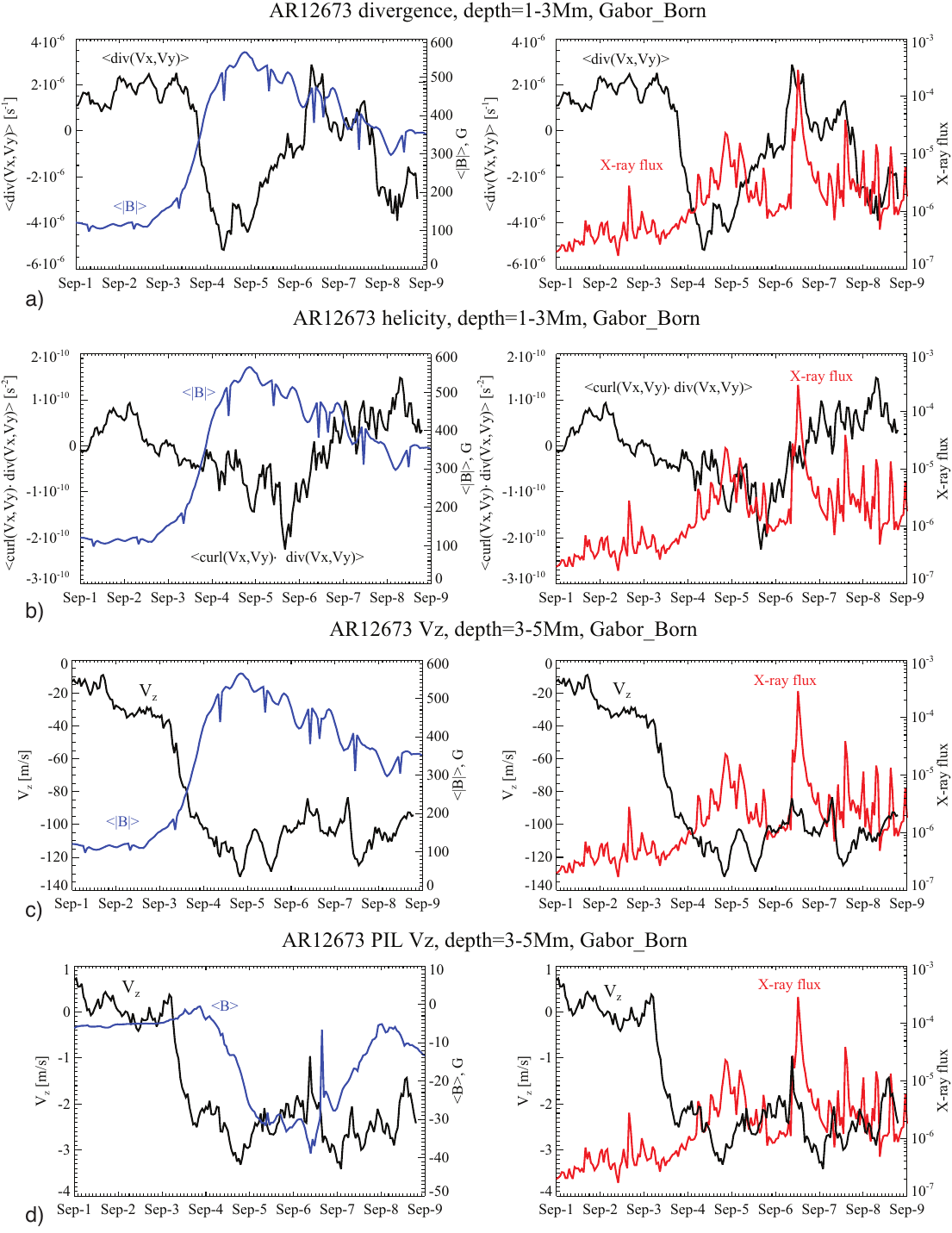} 
		\caption{Comparison of the mean magnetic field (left panels) and the flaring activity (right panels) of AR~12673 with subsurface flow characteristics: a) the mean horizontal divergence, $\left<{\rm div}(V_x,V_y)\right>$, averaged in the $1-3$~Mm deep layer; b) the mean kinetic helicity proxy, $\left<{\rm div}(V_x,V_y){\rm curl}(V_x,V_y)\right>$, in the same layer; c) the mean vertical velocity, $v_z$, in the $3-5$~Mm deep layer; and d) the vertical velocity averaged over the area around the polarity inversion line (near the center of the right panels in Fig.~\ref{fig2}). 
		}
		\label{fig7}
	\end{center}
\end{figure}

The comparison of the flow characteristics in the 1-3~Mm deep layer (mean divergence, helicity proxy, and vertical velocity component) with the mean magnetic flux, represented in the plots by mean unsigned magnetic field (Fig.~\ref{fig7}, left panels a) and b) shows the significant increase in the flow convergence (corresponding to the negative divergence) and an increase in the kinetic helicity. In addition, we observe the development of downflows beneath the active regions, illustrated for the 3-5~Mm deep layer in Fig.~\ref{fig7}, left panel c). 

The comparison of the flow characteristics with the soft X-ray flux (1-hour averages from the GOES satellite data) in Figure~\ref{fig7} (right panels) reveals a correlation of the large flares with enhancements of flow divergence. There is evidence of upflows in the vicinity of the polarity inversion line (Fig. \ref{fig7}, right panel d) prior to the big flares, X2.2 and X9.3 on Sept.6, and X1.3 on Sept.7, 2017.

\section{Correlation of Flow Characteristics with  Flare Productivity}
\begin{figure}[h!]
	\begin{center}
		\includegraphics[width=0.95\linewidth]{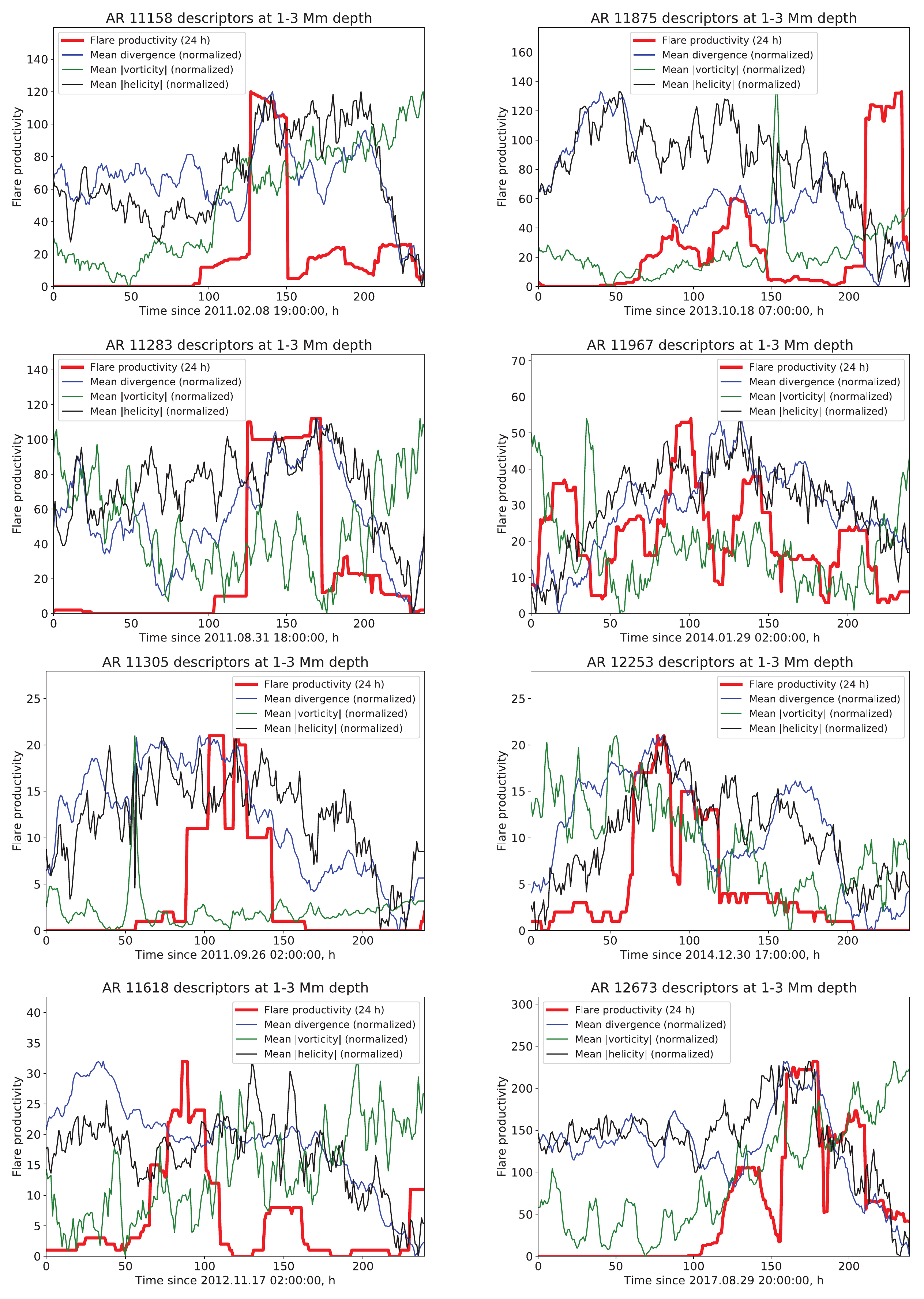} 
		\caption{Comparison of the flare productivity during the consecutive 24-hour periods (red lines) with the subsurface flow characteristics at the depth $1-3$~Mm: the mean horizontal divergence (blue), the mean vertical vorticity (green), and the mean helicity proxy (black), for the 8 active regions, listed in Table 1.
		}
		\label{fig8}
	\end{center}
\end{figure}

To investigate the potential of using the flow characteristics for the flare prediction, we performed a correlation analysis for eight active regions. We define the flare productivity of the active region as $P=N_{C}+10 \times N_{M}+100 \times$ $N_{X}$ where ${N}_{{C}}, {N}_{{M}}$, and ${N}_{{X}}$ are the total number of C, M, and X-class flares that were observed by the GOES satellite in the active regions within 24 hours from the considered moments. The flare productivity is used for correlation analysis with the active region magnetic and flow descriptors shown in Fig.~\ref{fig8}.

We analyzed correlations of the flow descriptors, the horizontal divergence, the vertical vorticity, and the helicity proxy, averaged over the whole $30\times 30$-degree regions tracked for 10-day during their passage on the solar disk,  with the flare productivity of the parental active region within the next 24-hour window. In addition to classically-used Pearson's correlation coefficients, which checks for linear dependence between the set of pairs of parameters, we analyze non-parametric Kendall's tau correlation coefficient defined as:
$$
\tau=\frac{2}{n(n-1)} \sum_{i<j} {\rm sgn}\left(x_{i}-x_{j}\right) {\rm sgn}\left(y_{i}-y_{j}\right)
$$
where the $x_{i}$ and $y_{i}$ are the values of the considered pair of parameters; sgn is a sign operator; ${n}$ is a number of elements in each data set. Kendall's tau ranges between -1 and 1, and its value is expected to be 0 for independent data sets.

\begin{table}[h]
\begin{center}	
\caption{Strongest Kendall's tau rank correlation coefficients and corresponding time lags (varying from 0~hr to 24~hr) found for subsurface flow map descriptors at 1-3 Mm depth and flare productivity. The correlations are obtained for 8 ARs tracked for 10 days with the 1-hour cadence. The entries in italics indicate cases where correlation coefficients were not statistically significant (the corresponding p-value is larger than 0.05).\\}

\begin{tabular}{|p{1.0in}|p{1.1in}|p{1.1in}|p{1.1in}|} 
		&&&\\[-3.5ex]\hline
	\textbf{Active Region} & \textbf{Total divergence\newline at 1-3 Mm} & \textbf{Total vorticity\newline at 1-3 Mm} & \textbf{Total kinetic helicity at 1-3 Mm} \\	&&&\\[-3ex]\hline
	11158 & 0.14 / 0 h & 0.53 / 7 h & 0.47 / 0 h  \\  &&&\\[-3ex]\hline
	11283 & 0.61 / 0 h & -0.27 / 18 h & 0.40 / 15 h \\&&&\\[-3ex] \hline
	11305 & 0.65 / 20 h & -0.25 / 23 h & 0.46 / 2 h \\&&&\\[-3ex] \hline 
	11618 & 0.30 / 24 h & -0.43 / 7 h & {\it 0.08 / 0 h} \\&&&\\[-3ex] \hline 
	11875 & -0.38 / 0 h & {\it 0.10 / 7 h} & -0.16 / 22 h \\&&&\\[-3ex] \hline 
	11967 & 0.15 / 6 h & 0.20 / 0 h & 0.48 / 11 h \\&&&\\[-3ex] \hline 
	12253 & 0.41 / 24 h & 0.54 / 24 h & 0.65 / 2 h \\&&&\\[-3ex] \hline 
	12673 & 0.14 / 19 h & 0.65 / 0 h & 0.22 / 20 h \\[-1ex] \hline 
\end{tabular}
\end{center}
\end{table}

Table 1 shows the strongest Kendall's tau rank correlation coefficients and corresponding time lags (varying from 0~hr to 24~hr) found for subsurface flow map descriptors at 1-3 Mm depth and flare productivity. These results show that, in some cases, a significant correlation can be found 24 hours prior to the flares. Thus, the subsurface flow characteristics even averaged over a large area around the active regions, may provide valuable information for predicting the flaring activity of active regions. These results warrant a more complete statistical analysis of the relationship between the subsurface flows and the flaring activity of active regions. 

\section{Discussion}

Detailed maps of subsurface flows inferred by time-distance helioseismology provide unique information about the subsurface dynamics during the emergence and formation of active regions and the periods of their flaring activity. The presented examples of emerging active regions show that the subsurface large-scale flows play a critical role in the active region formation and evolution. 
Each flow map was obtained by analyzing an 8-hour series of solar oscillations observed in the SDO/HMI data. To track the active region evolution, the flow maps were obtained with a one-hour cadence and ten subsurface layers in the top 19 Mm of the solar convection zone. 
The most significant variations of subsurface flows were observed in the top three layers in the depth range of 0--6~Mm, illustrated for five active regions in Figures~\ref{fig2}-\ref{fig6}. In these layers, we observed the formation of vortexes and shearing flows that contributed to the formation of the delta-type magnetic configurations, which were sources of powerful solar flares. The typical flow pattern during the initial emergence of bipolar magnetic regions consists of the flows diverging from the central area between the magnetic polarities observed in the line-of-sight magnetograms. A significant portion of the total magnetic flux can emerge in the central area between the forming sunspots, e.g. in the case of AR~13189, about half of the total flux emerged in such area in the form of small-scale bipolar elements, which moved to the large-scale magnetic structures (sunspots) of the corresponding polarities, contributing to their formation. This process continued for about 3 days. It means that the emerging magnetic flux represents a large-scale, highly fragmented structure and that the accumulation of magnetic field in sunspots is a result of magnetic self-organization in the near-surface of the Sun. The stability of sunspots is likely to be supported by the compact converging flows observed beneath the sunspots in the top 6~Mm deep layers. Outside the sunspots, we observe large-scale outflows extended to about 20~Mm outside the sunspot penumbrae.  The speed of these flows is about 100-200 m\,s$^{-1}$. In general, such flow pattern in the developed sunspots corresponds to the previous results of \citet{Kosovichev2000} and \citet{Zhao2001}. A similar flow structure was also recently found in the numerical simulations of emerging magnetic flux and sunspot formation by \citet{Hotta2020}.

Prior to and at the beginning of the flux emergence, the flow averaged over the whole tracked region is diverging. However, this flow becomes converging in the subsurface layers once a significant amount of the magnetic flux emerges. However, the large-scale flows are predominately divergent in the deeper layers. Such inflows and outflows around active regions were found in the earlier ring-diagram and time-distance helioseismology results \citep{Haber2004,Zhao2004}. 

It appears that, in addition to the subsurface flows caused by the interaction of the magnetic field with turbulent convection, the solar differential rotation may play a significant role. In particular, the newly emerging flux typically rotates faster than the already-formed sunspots. This leads to the interaction between the bipolar active regions and causes the formation of the flare-producing delta-type active regions. For example, examples, the delta-type AR 11158 was formed when the fast-moving leading positive polarity magnetic patch of the bipolar region, which emerged behind another bipolar region, collided with its trailing negative polarity spot, creating the flare-producing delta-type configuration. Active region 12673 presents an even more dramatic example of the interaction of an emerging bipolar magnetic region in the vicinity of a large sunspot.  A possible explanation of such interactions is that the emerging flux is anchored deeper in the near-surface shear layer than the developed sunspots and, thus, rotates faster due to the increasing rotation rate with the depth \citep{Getling2019a}. However, the subsurface flow maps showed that the fast eastward motion of the emerging magnetic flux was caused by the local flows associated with this flux. The physical mechanism driving these flows is not yet understood. 

The subsurface maps showed substantial shearing and converging flows in the areas beneath the polarity inversion lines of the delta-type sunspots. Similar shearing flows were observed on the surface by using various local correlation tracking and flow reconstruction techniques. Our results showed that, in addition to the shearing flows along the polarity inversion line, there were flows crossing this line. The subsurface flows might contribute to stressing the magnetic field in the polarity inversion line, intensifying the electric currents, and triggering the release of magnetic energy. In particular, a significant correlation was found between the flow divergence and helicity in the active regions with their flaring activity, although an extensive statistical analysis has not been performed.  

To evaluate the potential of the subsurface flow maps for flare prediction, we analyzed correlations of the flow descriptors (the total divergence, vorticity, and helicity averaged over the 30$\times$30-degree patches) with the flare productivity within the next 24-hour window for a sample of eight active regions.  The strongest Kendall's tau rank correlation coefficients and corresponding time lags (varying from 0~hr to 24~hr) were found for subsurface flow map descriptors at 1-3 Mm depth and flare productivity. These results indicate that measuring characteristics of subsurface flows can contribute to flare forecasting. However, more extensive statistical analysis and optimization of the flow characteristic calculations (e.g. defining spatial and temporal windows) are required.

In summary, the results show that the subsurface flow dynamics play a significant role in the formation, evolution, and flaring activity of active regions. The flow characteristics have the potential to improve the prediction of periods of flaring activity in active regions.\\

{\bf Acknowledgments}\\
The work was partially supported by NASA grants: NNX14AB70G, 80NSSC19K0268, 80NSSC19K0630, 80NSSC20K0602, 80NSSC20K1870, 80NSSC22M0162, 80NSSC23K0097, and  NSF grants 1936361 and 1835958. 

\bibliographystyle{iaulike}


\end{document}